**Noncoding RNAs evolutionarily extend animal lifespan**

Anyou Wang ORCID 0000-0002-4981-3606

Feinstone Center for Genomic Research, University of Memphis, Memphis, TN 38152, USA

Contact:

**anyou.wang@alumni.ucr.edu**

**Abstract:** The mechanisms underlying lifespan evolution in organisms have long been mysterious. However, recent studies have demonstrated that organisms evolutionarily gain noncoding RNAs (ncRNAs) that carry endogenous profound functions in higher organisms[1,2], including lifespan[3]. This study unveils ncRNAs as crucial drivers driving animal lifespan evolution. Species in the animal kingdom evolutionarily increase their ncRNA length in their genomes, coinciding with trimming mitochondrial genome length. This leads to lower energy consumption and ultimately lifespan extension. Notably, during lifespan extension, species exhibit a gradual acquisition of long-life ncRNA motifs while concurrently losing short-life motifs. These longevity-associated ncRNA motifs, such as GGTGCG, are particularly active in key tissues, including the endometrium, ovary, testis, and cerebral cortex. The activation of ncRNAs in the ovary and endometrium offers insights into why women generally exhibit longer lifespans than men. This groundbreaking discovery reveals the pivotal role of ncRNAs in driving lifespan evolution and provides a fundamental foundation for the study of longevity and aging.

## Introduction

It remains unclear how organisms evolve their lifespans[4–6]. For instance, humans can live over 100 years, but mice can only live < 5 years[4] although they share more than 80% of orthologous protein-coding genes[7]. This mystery has evolved into a fascinating topic that demands urgent attention since life expectancy has risen significantly over the past century.

Although lifespan is a complex and multifactorial process[6,8], it is fundamentally an evolutionary process in which genetic factors evolve to cope with lifespan evolution[5,8]. Thus it is essential to uncover these gene factors contributing to lifespan variations among different species. Current studies have paid much attention to protein-coding genes in the search for longevity determinants, but results from these studies have not provided sufficient evidence to explain the evolutionary lifespan disparity even between a small group of species or individuals[5,6]. Large scale lifespan differences between species remain elusive.

When species' genomes evolve, they usually gain a greater length of noncoding RNAs (ncRNAs) than proteins. For instance, the human genome contains a large number of ncRNAs, while most proteins remain similar[7]. The significance of these ncRNAs lies in the fact that they are actively transcribed and have their own functional system[1]. In addition, they execute important fundamental functions[2,9], including extending lifespans[3]. Therefore, ncRNAs may play a key role in organism lifespan evolution.

## Results

### Noncoding RNAs as evolutionary longevity drivers

To understand the evolutionary drivers of animal lifespan, we investigated genetic factors driving lifespan evolution across species in the animal kingdom. We collected both lifespan and genome data available today. Lifespan data were available for 4215 species samples[4], but only 333 have been sequenced with complete genome annotations (methods). These 333 samples were used for downstream analysis in this study.

Both Loess and linear regression were performed to examine the evolutionary association between the lifespan and length of protein-coding regions (referred to as proteins thereafter) and noncoding RNAs (ncRNAs) that are not protein-coding. Protein length was not correlated with lifespan (Figure 1A left). Furthermore, relative protein length (protein length/genome length, methods) is negatively correlated with lifespan (coefficient = -0.000614 pvalue=0.018, Figure 1A right), suggesting proteins as factors shorten lifespan. In contrast, absolute ncRNA length was significantly associated with lifespan (p<0.001 and coefficient 6.4*105, Figure 1B left), and relative ncRNAs also trended positively associated with lifespan (Figure 1B right), indicating that ncRNAs are evolutionarily associated with animal lifespan extension.

We further compared the genome evolution between human (lifespan >100 years) and mouse (<5 years). The human genome has evolved to carry many more ncRNAs than the mouse genome in terms of both total length (pvalue=0, chi-square test for this study, Figure 1C, left) and total number of genes (pvalue=3.646071e-123, Figure 1C, right). Human genome has more than 2.67 folds longer ncRNA length (650012775/243294640 bp) than mouse, while only a slight difference was observed in protein length (1379651157/1073396684 bp, Figure 1C left). In terms of gene number, the mouse genome has even more protein genes than the human genome (21908/20040, Figure 1C right). This consistently suggests that noncoding RNAs drive lifespan evolution.

**Noncoding RNAs evolutionarily associate with mitochondrial evolution**

Animal mitochondrial genome evolution is too slow to be associated with significant lifespan evolution (Figure 2A). We scaled down the observation to the mammalia class since this class had the most abundant samples. For each family in this class, we calculated the average mitochondrial genome length and the average length of genome proteins and ncRNAs. We then examined the correlation between them. Protein length was not correlated with mitochondrial genome length (Figure 2B), but ncRNA length trended to be negatively correlated with mitochondrial genome length (Figure 2C) and the relative ncRNA length was significantly correlated with mitochondrial genome length (Figure 2D), suggesting that ncRNAs are evolutionarily associated with trimming the mitochondrial genome.

**Long-life ncRNA motifs**

To identify long-life motifs, we first generated a total of 22868 possible motifs from the complete permutation of 4 bases (ATCG) from 1 to 7 bases ($4^1+4^2+4^3+4^4+4^5+4^6+4^7$), and then calculated the frequency for each motif (methods), which measures the number of a ncRNA motif occurring in each sample. The motif frequency of 333 samples was constructed into a matrix used to find the most significant motifs as long-life motifs. This was done via the FINET[10] computational algorithm (methods). FINET set the lifespan as a target and found significant motifs regulating the target (methods).

With FINET frequency score > 0.6, 4 motifs were selected (Figure 3A), including two long-life motifs with coefficient >0 and two short-life motifs with coefficient <0. These two long-life motifs, GGTGCG and CGTATA, evolve in a gradient-gain pattern along with increasing lifespan across species, while the 2 short-life motifs, ACGTCG and TCTCTC, evolutionarily lose their contents when lifespan evolves longer (Figure 3B). These 2 long-life motifs were significantly associated with lifespan extension (Figure 3C-3D). In contrast, the short-life motif did not significantly associate with lifespan (Figure 3E). These together suggested the two long-life motifs, GGTGCG and CGTATA as the key motifs driving animal lifespan evolution.

**Long-life motifs drive human evolution**

We further examined the evolutionary pattern of these ncRNA motifs in humans and mice. These two long-life motifs, GGTGCG and CGTATA, were significantly higher in the noncoding regions that remove the protein-coding regions in the human genome than in mice (Pvalue = 2.726858e-60, chisq.test, Figure 4A), but the short-life motif, TCTCTC, was much higher in mice than in man (pvalue = 1.062907e-69, Figure 4A). This suggested that the human genome evolved to increases the contents of long-life motifs while decreasing that of short-life motifs.

The above pattern was further demonstrated by the chromosome distribution of GGTGCG (Figure 4B), where human frequency reached over 2000 versus 400 in mice. Moreover, the frequency distribution showed a gradient increasing pattern from chromosome Y to chromosome 2 in human, while this distribution pattern was relatively flat in mice, suggesting that human chromosomes actively evolve to cope with lifespan evolution.

The top 3 human ncRNA genes that contain the most GGTGCG are CASC15, LINC02934, and ENSG00000286481 (Figure 4C). CASC15 contains 69 GGTGCG motifs (Figure 4C), suggesting CASC15 as a long-life gene. We examined its expression in 27 normal tissues and 171 RNAseq samples[11] (methods) and found that CASC15 was highly expressed in three tissues, including the endometrium, the ovary, and the cerebral cortex. The remaining two ncRNA genes, LINC02934, and ENSG00000286481, are highly expressed in the testis. This suggests the reproductive and nervous system as the most critical system for human lifespan.

It was found that only two mouse ncRNA genes had more than 20 motifs, Snhg14 and 9530036011Rik (Figure 4D), both highly expressed in the cerebral cortex out of 16 tissues and 1220 RNAseq samples[12] (methods), which the cerebral cortex is the key tissue driving the evolution of mouse lifespan.

## Discussion

This study uncovered noncoding RNAs as evolutionary drivers of animal lifespan. NcRNA evolution coincides with the mitochondrial genome trimming. Shortening the mitochondrial genome reduces energy consumption and extends lifespan. Homeostasis states and lifespan are determined by energy consumption. High energy consumption shortens lifespan. Protein synthesis and function consume more energy than ncRNAs[3], so it is reasonable for ncRNAs to drive organism lifespan evolution, instead of proteins as thought.

Noncoding RNA motifs drive lifespan evolution. Long-life motif contents increase with lifespan extension, while short-life motif contents decrease. Long-life motifs concentrate on the nervous system and reproductive system, suggesting that these two systems are paramount to lifespan evolution. Furthermore, genes containing these motifs, such as CASC15, LINC02934, and ENSG00000286481, are crucial ncRNAs driving lifespan evolution although their functions have not been fully uncovered. Especially, CASC15 is a strong driver of long lifespan even though it is defined as cancer susceptibility 15. These functions should be redefined.

Both humans and mice evolve in the cerebral cortex for lifespan. Humans, however, have also evolved a highly developed reproductive system that extends lifespan, whereas mice do not. The difference in their reproductive systems explains the lifespan gap between them.

This reproductive system difference further explains the lifespan disparity between women and men. Women live longer than men[13]. Current literature explains this disparity through a series of factors, such as the X and Y chromosomes, different levels of infection, disease, and environmental factors[14–17]. These only apply to small groups of people or a human community, but not at the population level. At the population level, these factors can be normalized as noise and ignored. The fact is that women suffer more serious diseases than men. For example, breast cancer kills more than 685,000 of women every year[18]. Fundamental factors should be evolutionary factors responsible for lifespan disparity, but they remain uncovered. Here, we found that women evolutionarily carry ovaries and endometriums with more longevity motif contents than men who carry testis. The reproductive system of women is an evolutionary factor contributing to women's longer longevity.  In conjunction with this study, it was discovered that ovary lifetime is associated with female longevity and that removing the ovary shortens female life expectancy[19]. Additionally, our finding that reproductive systems determine lifespan gaps is also supported by the recent discovery showing that prepubertal casting eliminates the sex gap between men and women[20].

Overall, this study provides a fundamental foundation for studying longevity and aging. They pave the way for the development of novel therapeutic strategies aimed at promoting healthy aging and extending lifespan in humans.

**Methods**

See supplemental materials

**Declarations**

Ethics approval and consent to participate

NA

Consent for publication

NA

Availability of data and material

NA

Competing interests

NO

Funding

No funding resource associated with this project

Authors' contributions

All

**Figure legends**

Figure 1. Noncoding RNAs were evolutionarily associated with animal longevity. The correlations between longevity and noncoding RNA and protein length were examined by both Loess (green) and linear regression (red). A, Absolute protein length is not correlative to longevity (left), and relative protein length (length/genome length) is negatively correlative to lifespan (right). B, noncoding RNA length is significantly positive to lifespan (left) and relative noncoding RNAs do not correlate to lifespan. C, humans carry much more noncoding RNAs than mice in both length and number.

Figure 2. Noncoding RNAs are evolutionarily associated with mitochondrial genome length. Loess (green) and Linear (red) regression plots. A, lifespan does not significantly relate to mitochondrial genome size, although there is a negative trend between them. B, the average protein lengths of mammal families do not correlate with their mitochondrial genome lengths. C, the average noncoding lengths of mammal families trend to negatively correlate with their mitochondrial genome length. D, negative correlation between mean relative noncoding lengths and mitochondrial genome length among mammal families.

Figure 3. Motifs drive longevity evolution. A, Top 4 motifs derived from FINET. B, Top 4 heatmap of motif frequency across 334 species along with lifespan. Row as species and column as motif. Motif frequency increases from blue to red. The row was ordered by lifespan gradient as shown in the separate right column. C-D, Regressions between lifespan and motifs. Zscore was calculated for motif frequency and lifespan from B-D.

Figure 4. Motifs drive human and mouse evolution. A, The frequency of 4 motifs in humans and mice. B, GGTGCG motif distribution in humans (left) and mice (right). C, top 3 human genes with the highest GGTGCG contents (left) and the 3 tissues with the highest expression of these top 3 genes. D, top 2 mouse genes with the highest GGTGCG contents (left) and the 3 tissues with the highest expression of these top 2 genes.

Figure 1

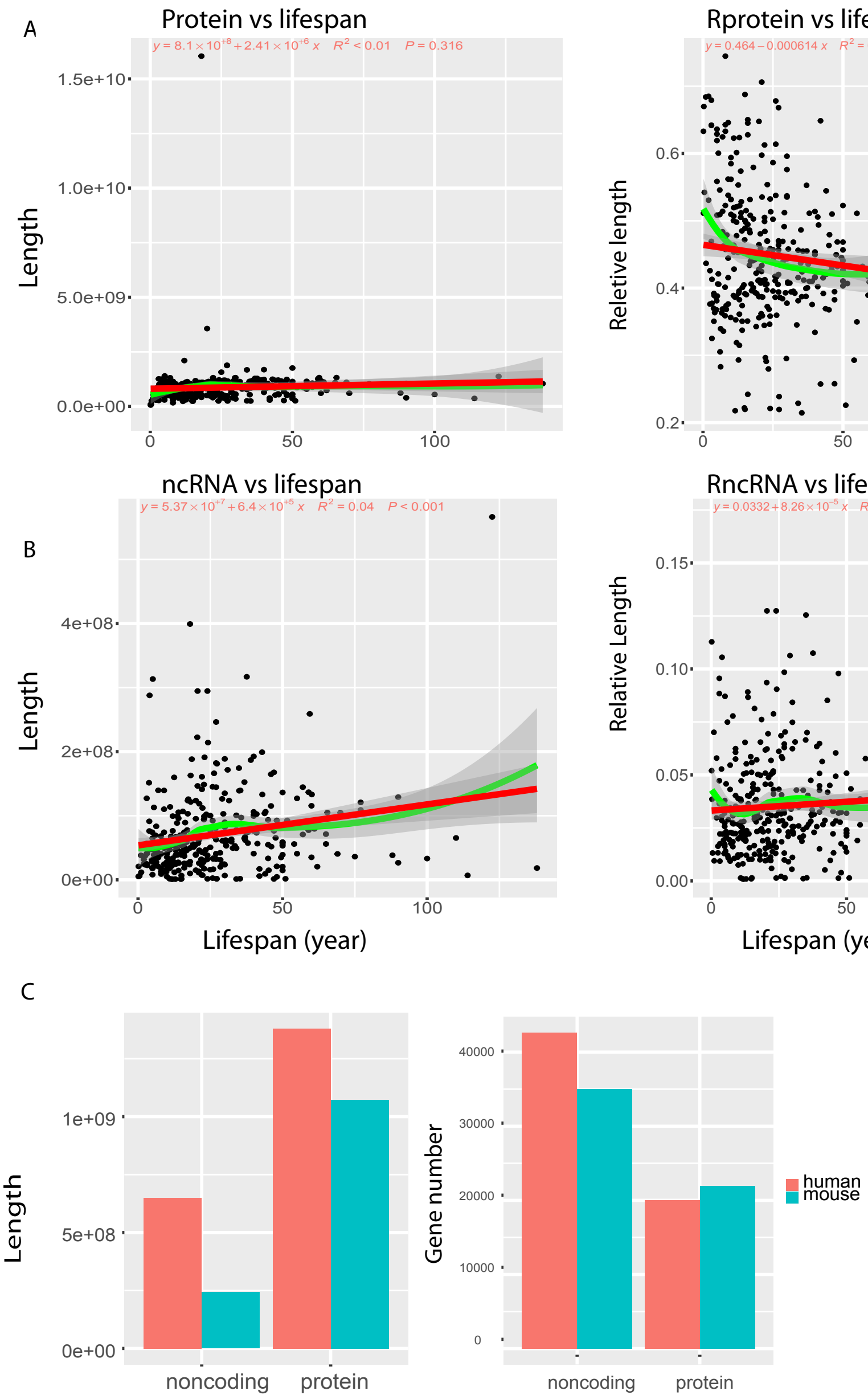

Figure 2

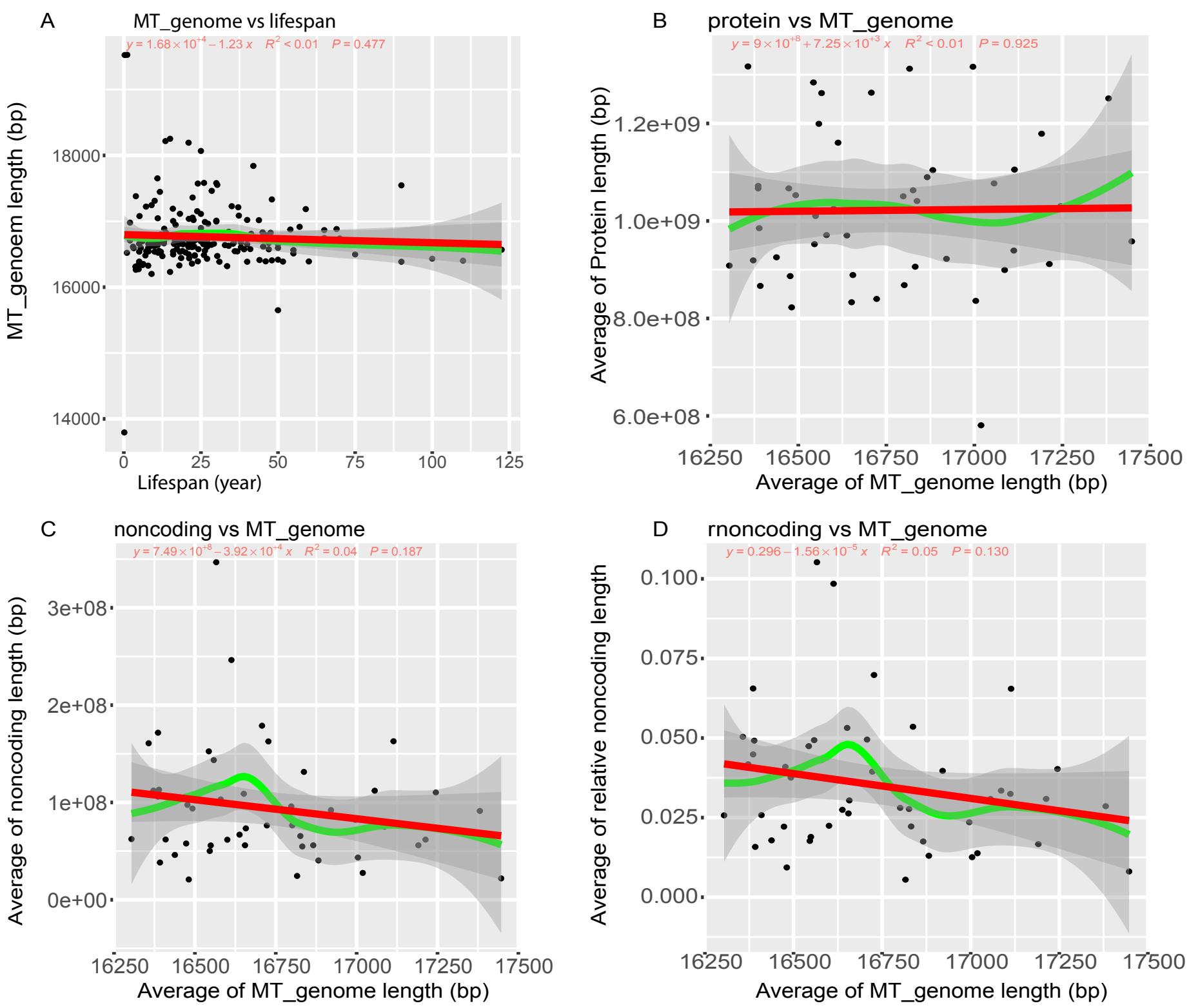

A
MT_genome vs lifespan
$y = 1.68 \times 10^{+4} - 1.23\,x \quad R^2 < 0.01 \quad P = 0.477$
MT_genoem length (bp)
14000
16000
18000
0
25
50
75
100
125
Lifespan (year)
B
protein vs MT_genome
$y = 9 \times 10^{+8} + 7.25 \times 10^{+3}\,x \quad R^2 < 0.01 \quad P = 0.925$
Average of Protein length (bp)
6.0e+08
8.0e+08
1.0e+09
1.2e+09
16250
16500
16750
17000
17250
17500
Average of MT_genome length (bp)
C
noncoding vs MT_genome
$y = 7.49 \times 10^{+8} - 3.92 \times 10^{+4}\,x \quad R^2 = 0.04 \quad P = 0.187$
Average of noncoding length (bp)
0e+00
1e+08
2e+08
3e+08
16250
16500
16750
17000
17250
17500
Average of MT_genome length (bp)
D
rnoncoding vs MT_genome
$y = 0.296 - 1.56 \times 10^{-5}\,x \quad R^2 = 0.05 \quad P = 0.130$
Average of relative noncoding length
0.000
0.025
0.050
0.075
0.100
16250
16500
16750
17000
17250
17500
Average of MT_genome length (bp)

Figure 3

A

| Motif | Frequency | Coefficient |
|---|---|---|
| ACGTCG | 0.675 | -0.08885553 |
| GGTGCG | 0.66 | 0.1645365 |
| TCTCTC | 0.655 | -0.05143238 |
| CGTATA | 0.625 | 0.08574241 |

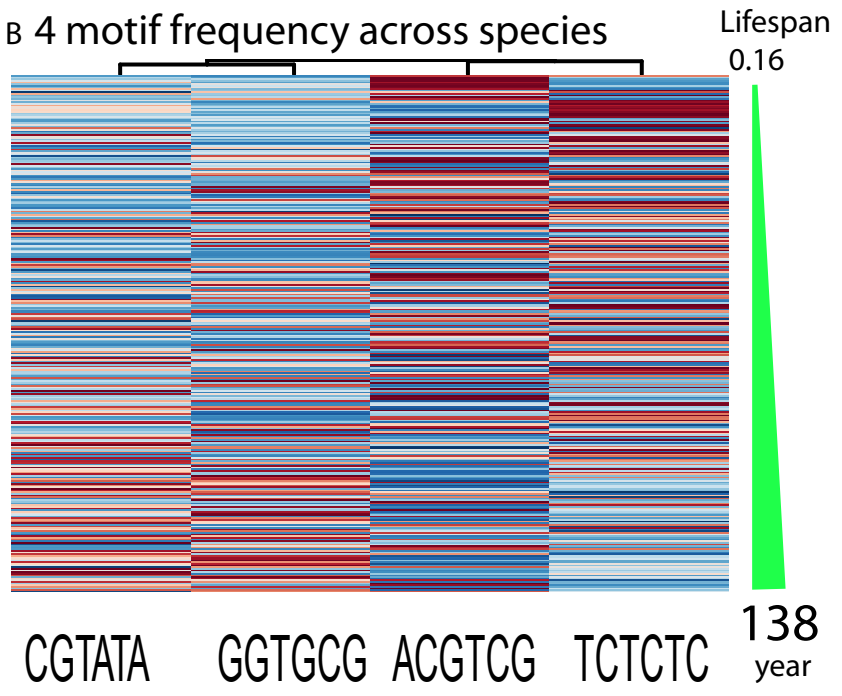

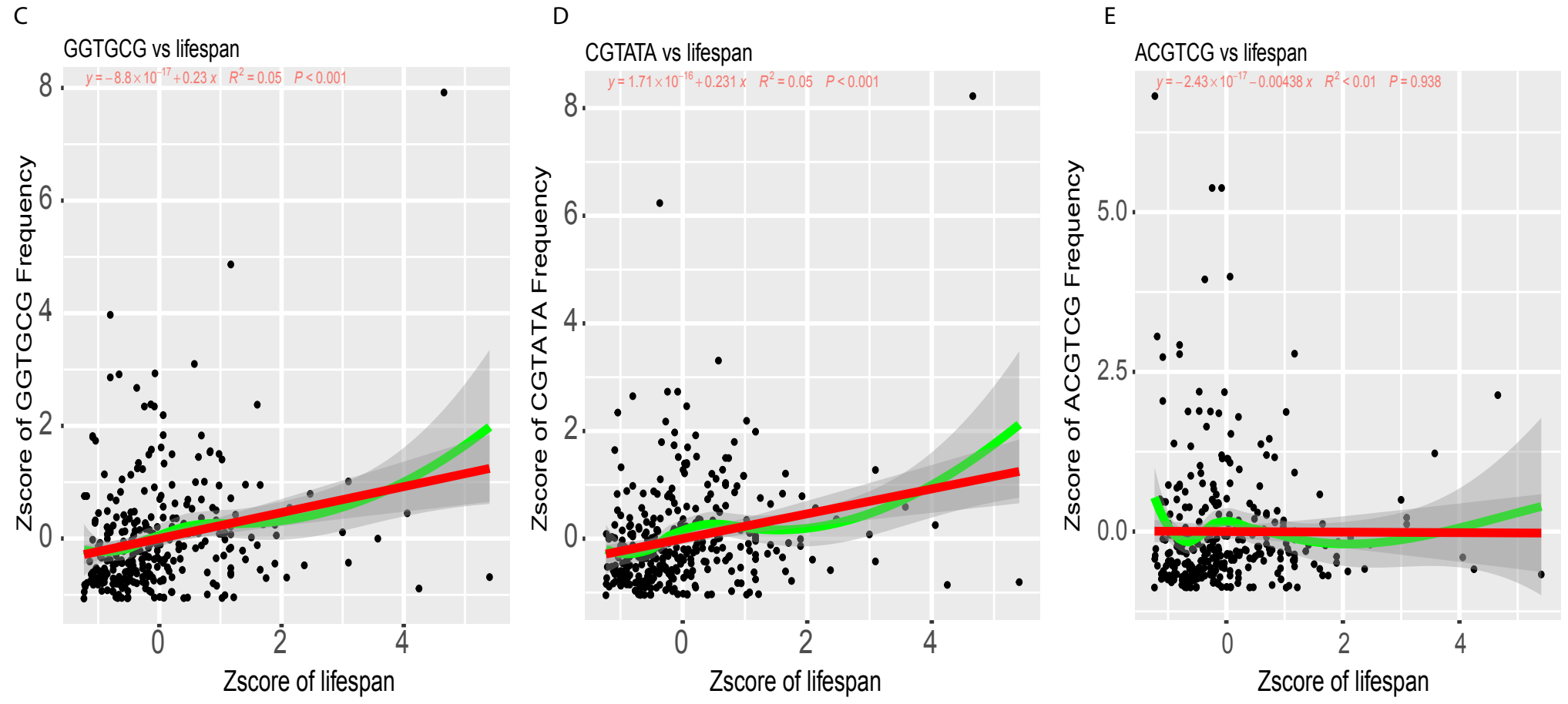

Figure 4

A

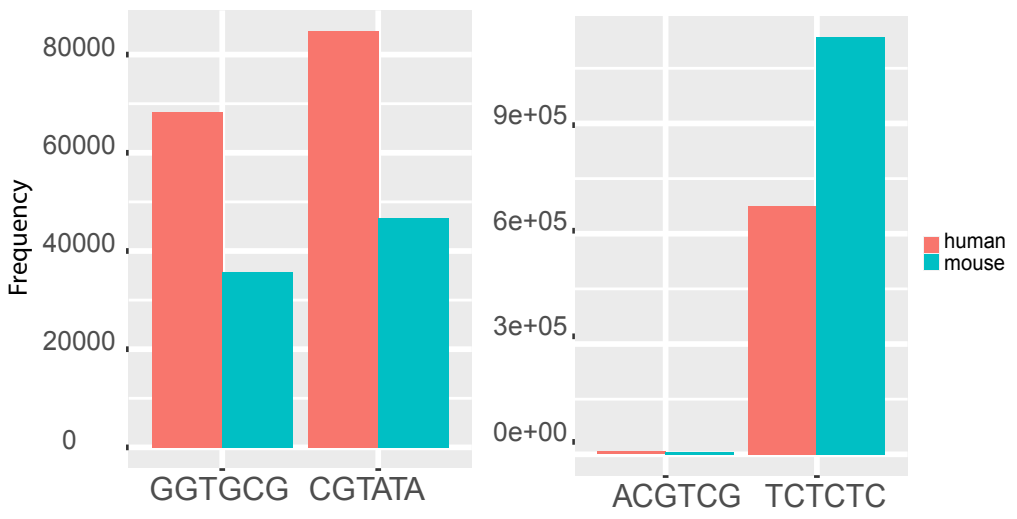

B

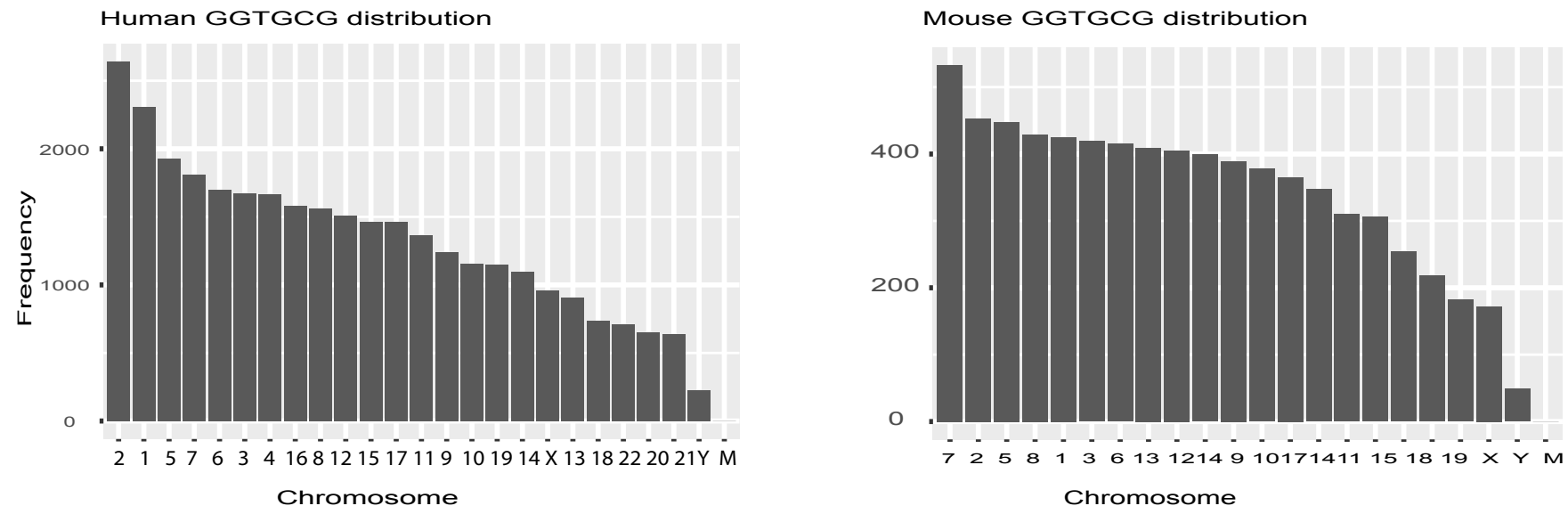

C

Top 3 human genes

| ID | symbol | Frequency |
|---|---|---|
| ENSG00000272168.10 | CASC15 | 69 |
| ENSG00000204929.13 | LINC02934 | 43 |
| ENSG00000286481.2 | ENSG00000286481 | 43 |

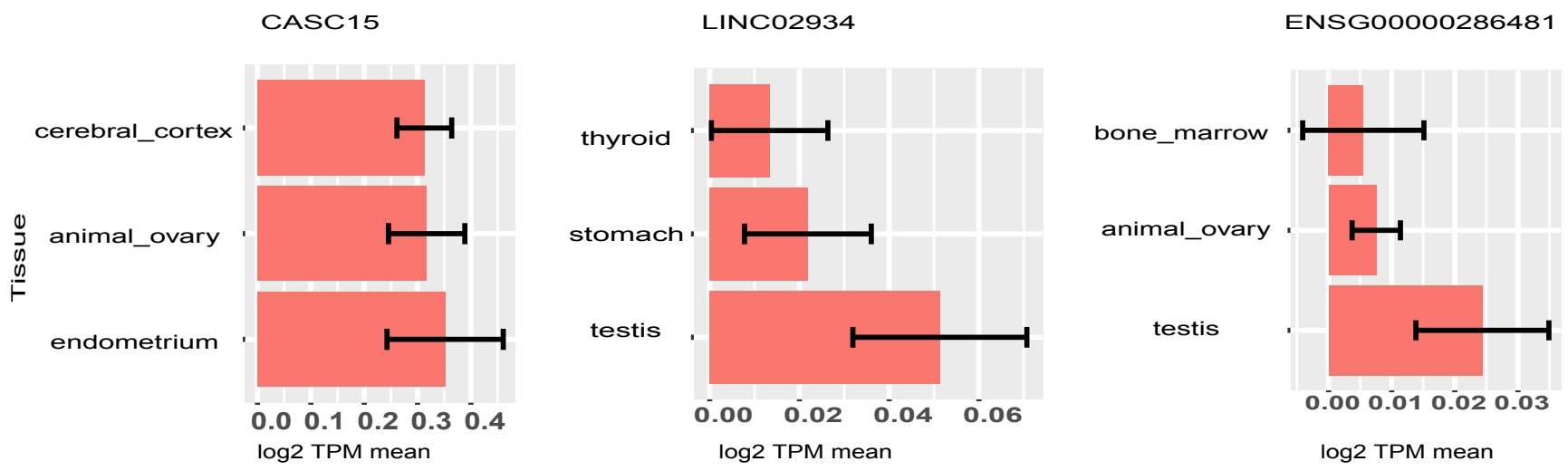

D

Top 2 mouse genes

| ID | symbol | Frequency |
|---|---|---|
| ENSMUSG00000100826.7 | Snhg14 | 25 |
| ENSMUSG00000097726.8 | 9530036O11Rik | 22 |

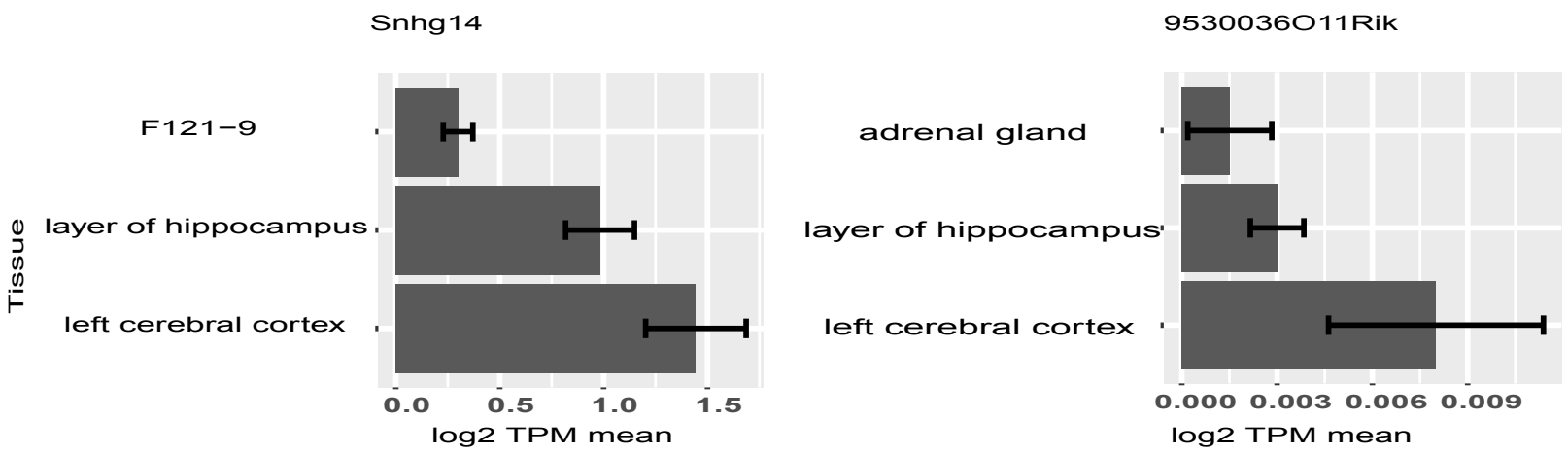